\documentclass[12pt,preprint]{aastex}
\begin{document}
\title{Formation of ultra-compact X-ray binaries through
circum-binary disk-driven mass transfer}
\author{Bo Ma and Xiang-Dong Li}
\affil{Department of Astronomy, Nanjing University, Nanjing 210093,
China;\\ xiaomabo@gmail.com, lixd@nju.edu.cn}

\begin{abstract}
The formation of ultra-compact X-ray binaries (UCXBs) has not been
well understood. Previous works show that ultra-short orbital
periods ($<1$ hr) may be reached through mass transfer driven by
magnetic braking in normal low/intermediate-mass X-ray binaries
(L/IMXBs) only for an extremely small range of initial binary
parameters, which makes it difficult to account for the rather large
population of UCXBs. In this paper we report the calculated results
on mass transfer processes in L/IMXBs with a circum-binary disk. We
show that when the orbital angular momentum loss due to a
circum-binary disk is included, ultra-short orbital periods could be
reached for a relatively wide range of initial binary parameters.
The results of our binary models suggest an alternative formation
channel  for  UCXBs.
%We also discuss the possible initial system parameters and current
%evolutionary state of one typical
\end{abstract}


\section{Introduction}
Ultra-compact X-ray binaries (UCXBs) are X-ray sources with very
short orbital periods ($P<1$ hr). They are thought to be powered by
accretion from a Roche-lobe (RL) filling donor star to a neutron
star (NS). The donor has to be compact, such as the helium (or more
heavier elements) core of an evolved giant star or a white dwarf
(WD), to fit in the small RL \citep{nelson86}. Spectra observations
have shown possible C/O or He/N features in some of the UCXBs or
UCXB candidates \citep[][and references
therein]{nelemans04,nelemans06}. Particular interests have been paid
to UCXBs in recent years since they are thought to be potential
sources for the gravitational-wave detector {\em LISA} \citep{nelemans09}.

Scenarios for the formation of the UCXBs can be summarized as follows.
For UCXBs in globular clusters, they often
invoke dynamical processes, such as (1) direct collisions between a
NS and a giant \citep{verbunt87,davies92,
rasio91,ivanova05,lombardi06}, (2) tidal capture of a low-mass
main-sequence (MS) star by a NS \citep{bailyn87}, and (3) exchange
interaction between a NS and a primordial binary
\citep{davies98,rasio00}. All these dynamical scenarios are involved
with the so-called common-envelop (CE) phase when the mass transfer
between a NS and a (sub)giant is dynamically unstable, which may
help form a tight NS$+$WD or NS$+$He star binary. After that, the
orbital period decays to ultra-short regime through gravitational
radiation (GR) until the WD/He star overflows its RL. For UCXBs in the
Galactic field where dynamical
collisions are not important, generally two CE phases are required to
form a tight NS$+$WD or NS$+$He
star binary, which decays to the ultra-short regime through GR
\citep{tutukov93,iben95,yungelson02,belcz04}.
%The first dynamical scenario may result directly
%in a NS+WD or NS+helium core of a (sub) giant, while the second and third scenarios
%result in a NS+MS binary, then evolve to NS+WD through common envelop phase caused by unstable
%Roche lobe overflow mass trasfer between a NS+(sub) giant.
%An expanding giant star can engulf its NS companion
%and then evolve into a CE phase. After the CE phase, a compact binary of NS
%and the helium core of the giant will form and then evolve into a UCXB
%\citep{davies98,rasio00}. \citet{lombardi06} pointed out that direct collisions
%between NS and sub giant or small red giant could also produce NS+WD binaries
%which could evolve to UCXBs.
An alternative formation channel of UCXBs is through stable mass transfer in normal
low/intermediate-mass X-ray binaries (L/IMXBs) driven by magnetic
braking \citep[MB;][]{verbunt81}, which is now called the ``magnetic
capture" scenario \citep{paczynski81,pylyser88,pod02,nelson03,
sluys05a}. However, \citet{sluys05a} found that only a very small
range of initial parameters are allowed for the binaries to evolve
to UCXBs through this channel within the Hubble time. This makes it
impossible to account for the relatively large number of observed
UCXBs. Another problem of this scenario is related to the efficiency
of MB. Observations of some rapid rotators in young, open clusters
seem to be contradicted with the MB law originally suggested by
\citet{verbunt81}, and a modified magnetic braking law was proposed
to resolve this problem \citep{queloz98,sills00}. \citet{sluys05b}
showed that, if this new magnetic braking law is adopted, no UCXBs can
be formed through the magnetic capture channel.

In this paper we propose an alternative scenario for the formation of UCXBs
through mass transfer between a NS and a MS star. We assume that
during Roche-lobe overflow (RLOF), a small fraction of the mass lost
from the donor forms a circum-binary (CB) disk around the binary
rather accretes onto the NS \citep{heuvel94}. Previous works have
shown that a CB disk can extract orbital angular momentum from the
binary effectively \citep{spruit01,taam01}, and enhance the mass
transfer rate, thus considerably influence the binary evolution
\citep{chen06,chen06b}. In this work we explore the possible role of
CB disks in the formation of UCXBs. In \S 2 we introduce the binary
evolution code and the input physics incorporated in our
calculations. In \S 3 we present the calculated results  and compare
them with observations. The implications of our model and related
uncertainties are discussed in \S 4.

\section{Evolutionary code and Input physics}
\subsection{The stellar evolution code}
 We use the STAR programme, originally developed by \citet{eggleton71,eggleton72}
and updated by other authors \citep{han94,pols95}, to compute the
binary evolution. The ratio of mixing length to pressure scale
height $\chi=l/H_p$ is set to be 2.0 and convective overshooting
parameter to be $os=0.12$, implying a $0.24H_p$ overshooting
distance. The opacity table is taken from \citet{hubbard69},
\citet{rogers92} and \citet{alexander94}. Solar compositions
($X=0.70$, $Y=0.28$ and $Z=0.02$) are adopted. The binary system is
initially composed of a NS primary of mass $M_1$ and a
zero-age main-sequence (ZAMS) secondary of mass $M_2$
with an orbital period $P_{\rm i}$. The effective radius of the RL for
the secondary star is calculated from \citet{eggleton83} equation,
\begin{equation}
R_{\rm{L},2}=\frac{0.49q^{-2/3}a}{0.6q^{-2/3}+\ln(1+q^{-1/3})},
\end{equation}
where $q=M_2/M_1$ is the mass ratio of the binary components, and $a$
is the orbital separation. The rate of mass transfer via RLOF is
calculated with $-\dot{M}_2=RMT\cdot
\max[0,(R_2/R_{\rm{L},2}-1)^3]M_{\sun}$yr$^{-1}$ in the code, where
$R_2$ is the radius of the secondary, and we set $RMT=10^3$.

\subsection{Angular momentum loss mechanisms}
We assume that the secondary star rotates synchronously with the
binary orbital revolution, since the timescale of tidal
synchronization is generally much shorter than the characteristic evolutionary
timescale of the binaries considered here. We take into account four kinds of angular
momentum loss mechanisms described as follows.

\noindent (1) Gravitational radiation, which becomes important when
the orbital period is short. The angular momentum loss rate is given by
\citep{landau}
\begin{equation}
\frac{{\rm d}J}{{\rm d}t}|_{\rm GR}=-\frac{32}{5}\frac{G^{7/2}}{c^5}
\frac{M_1^2M_2^2(M_1+M_2)^{1/2}}{a^{7/2}},
\end{equation}
where $J$, $G$ and $c$ are the orbital angular momentum,
gravitational constant and speed of light, respectively.

\noindent (2) CB disk. We assume that a small fraction $\delta (\ll
1)$ of the mass lost from the donor feeds into the CB disk rather
leaves the binary, which yields a mass injection rate of the CB disk
as $\dot{M}_{\rm CB}=-\delta\dot{M_2}$. Tidal torques are then
exerted on the binary by the CB disk via gravitational interaction,
thus extracting the angular momentum from the binary system. The
angular momentum loss rate via the CB disk is estimated to be
\citep{spruit01}
\begin{equation} \label{cb}
\frac{{\rm d}J}{{\rm d} t}|_{\rm CB}=-\gamma\left(\frac{2\pi
a^2}{P}\right) \dot{M}_{\rm CB}\left(\frac{t}{t_{\rm
vi}}\right)^{1/3},
\end{equation}
where $\gamma^{2}=r_{\rm i}/a=1.7$ ($r_{\rm i}$ is the inner radius
of the CB disk), $t$ is the time since mass transfer begins. In the
standard $\alpha$-viscosity disk \citep{shakura73}, the viscous
timescale $t_{\rm vi}$ at the inner edge $r_{\rm i}$ of the CB disk
is given by $ t_{\rm vi}=2\gamma^{3}P/3\pi\alpha\beta^{2}$, where
$\alpha$ is the viscosity parameter (we set $\alpha=0.01$ in the
following calculations), $\beta=H_{\rm i}/r_{\rm i}\sim 0.03$
\citep{belle04}, and $H_{\rm i}$ is the scale height of the disk.
When $\dot{M}_{\rm CB}$ is sufficiently large, the mass transfer
will become dynamically unstable, so we add an ad hoc term
\[\exp(1+\dot{M}_2/2\dot{M}_{\rm Edd}) \;{\rm if}\; -\dot{M}_2>2\dot{M}_{\rm Edd}\]
to Eq.~(\ref{cb}) to suppress the CB disk-induced angular momentum
loss  rate when the mass loss rate is high. We find that this term
does not influence the evolutionary tracks considerably, only
expanding the parameter space of $\delta$ suitable for UCXB
formation within the Hubble time by $\sim 10\%-20\%$. Here the Eddington
accretion limit is expressed as
\begin{equation} \label{edd}
\dot{M}_{\rm
Edd}=3.6\times10^{-8}(\frac{M_1}{1.4M_\sun})(\frac{0.1}{\eta})
(\frac{1.7}{1+X})M_\sun {\rm yr}^{-1},
\end{equation}
where $\eta=GM_1/Rc^2$ is the energy release efficiency through
accretion ($R$ the NS radius), and $X$ is the mass fraction of
hydrogen in the accreting material.

\noindent (3) Mass loss. Similar as in \citet{pod02}, we assume that
the NS accretion rate is limited to the Eddington accretion rate,
and that when the mass transfer rate is less than $\dot{M}_{\rm
Edd}$,  half of the mass is accreted by the NS, i.e., $\dot{M}_1=
\min (\dot{M}_{\rm Edd}, -\dot{M}_2/2)$. The excess mass is
lost in the vicinity of the NS through isotropic winds, carrying
away the specific angular momentum of the NS, i.e.
\begin{equation} \label{eq:massloss}
 \frac{{\rm d} J}{{\rm d}t}|_{\rm ML} \simeq  \left\{
 \begin{array}{lll}
\frac{1}{2}\dot{M}_2 a_{1}^2\omega, & & |\dot{M}_2| < 2\dot{M}_{\rm Edd}\\
(\dot{M}_2+\dot{M}_{\rm Edd}) a_{1}^2\omega, & & |\dot{M}_2 |\ge 2\dot{M}_{\rm Edd}
\end{array}\right.
\end{equation}
where $a_1=aM_2/(M_1+M_2)$ is the orbital radius of the NS,  and
$\omega$ is the orbital angular  velocity of the binary.

\noindent (4) Magnetic braking. We use the saturated magnetic
braking law suggested in \citet{sills00},
\begin{equation} \label{eq:magneticbraking}
 \frac{{\rm d} J}{{\rm d}t}|_{\rm MB} =  \left\{
 \begin{array}{lll}
-K \omega^3 \left(\frac{R_2}{R_{\sun}}\right)^{1/2}
\left(\frac{M_2}{M_{\sun}}\right)^{-1/2},  & & \omega \leq \omega_{\rm{cr}} \\
 -K \omega_{\rm{cr}}^2 \omega
\left(\frac{R_2}{R_{\sun}}\right)^{1/2}
\left(\frac{M_2}{M_{\sun}}\right)^{-1/2}, & & \omega >
\omega_{\rm{cr}}
\end{array}\right.
\end{equation}
where $K=2.7\times10^{47}$ gcm$^2$s \citep{andronov03},
$\omega_{\rm{cr}}$ is the critical angular velocity at which the
angular momentum loss rate reaches a saturated state, and can be
estimated as \citep{krishnamurthi97},
\begin{equation}
\omega_{\rm{cr}}(t) = \omega_{\rm{cr},\sun}
\frac{\tau_{\rm{t}_0,\sun}}{\tau_{\rm{t}}},
\end{equation}
where $\omega_{\rm{cr},\sun}=2.9\times 10^{-5}$ Hz,
$\tau_{\rm{t}_0,\sun}$ is the global turnover timescale for the
convective envelope of the Sun at its current age, $\tau_{\rm{t}}$
for the secondary at age $t$, solved by integrating the inverse
local convective velocity over the entire surface convective
envelope \citep{kim96}. Following \citet{pod02}, an ad hoc factor is
also added to Eq.~(\ref{eq:magneticbraking})
\[
\exp(-0.02/q_{\rm{con}}+1) \; {\rm if}\;q_{\rm{con}}<0.02,
\]
where $q_{\rm{con}}$ is the mass fraction of the surface convective
envelope. This term is used to reduce the strength of MB when the
star has a very small convective envelope and hence does not have a
strong magnetic field.

\section{Results}
\subsection{Parameter space of $P_{i}$ and $M_{\rm 2,i}$}
\begin{figure}
\epsscale{0.80} \plotone{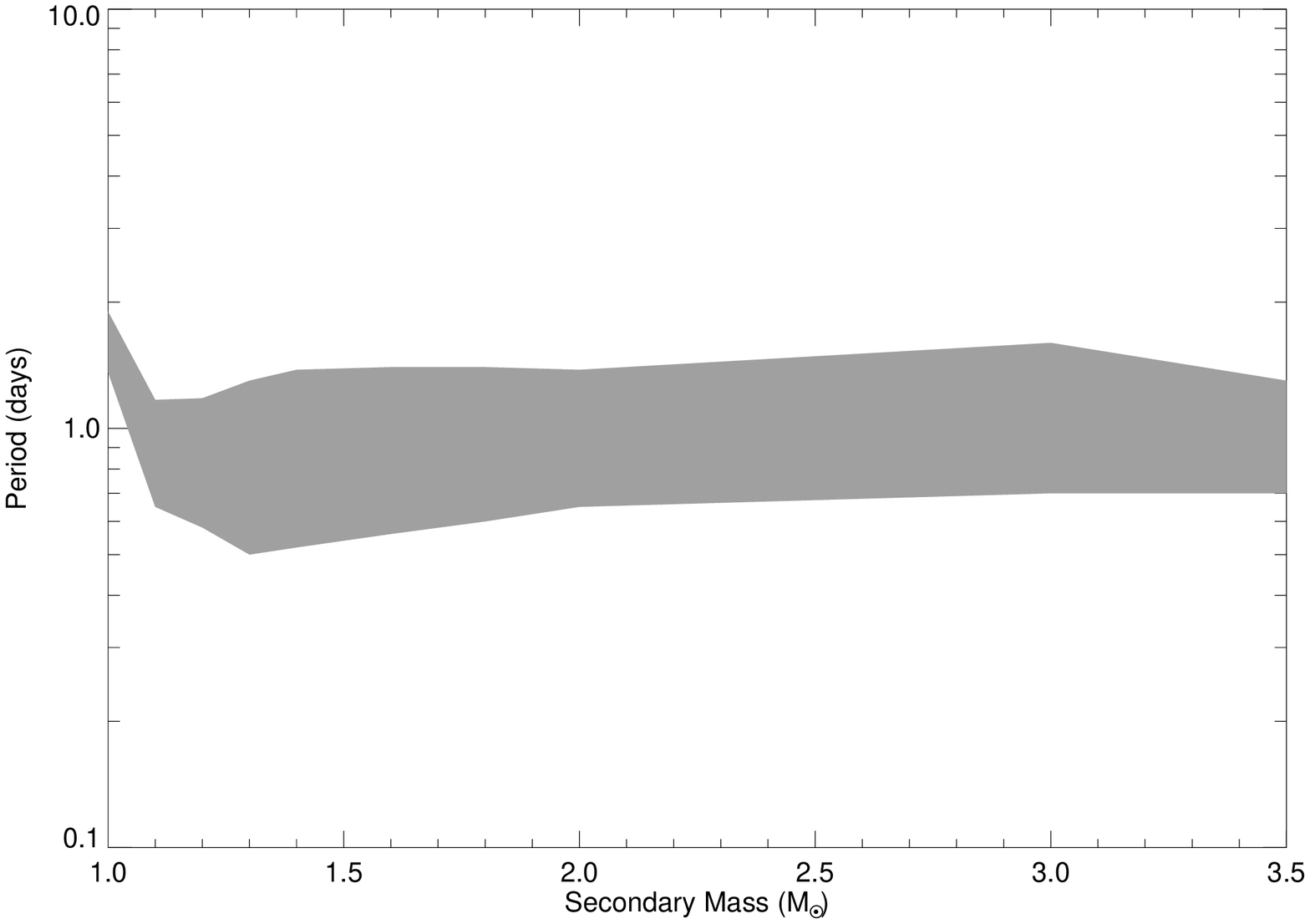} \caption{ Parameter space of the
initial orbital period and donor star mass for binary systems which
are able to evolve to UCXBs under the influence of a CB disk within 13.7
Gyr. \label{f1}}
\end{figure}

Similar as in \citet{sluys05a}, we define UCXBs as X-ray binaries with $P<50$
min. The binary systems we considered are initially composed of a
1.4 $M_\sun$ NS and a $0.5-5M_\sun$ ZAMS star. We have performed
calculations of a large number of binary evolutions, to search
suitable values in the three dimensional binary parameter space
$\delta$ (assumed to be constant through one evolutionary sequence), $P_{\rm i}$ and
$M_{\rm 2,i}$ for binaries evolved to UCXBs within the age of the
universe ($13.7$ Gyr). The distribution of $P_{\rm i}$ and $M_{\rm
2,i}$ for successful systems is shown in Fig.~\ref{f1} with $P_{\rm
i}\sim 0.7-1$ d and $M_{\rm 2,i}\sim 1-3.5\,M_{\sun}$. When the
donor mass $M_{\rm 2,i}> 3.5M_\sun$, the mass transfer becomes
dynamically unstable \citep[see also][]{pod02}. Comparing with the
results of \citet{sluys05b} one can find that there is a relatively
large parameter space for the formation of UCXBs if there is a CB
disk at work.

%Next is to discuss the M_2 range from dynamical unstable
%Tauris & savonije(1999) found that all systems with (sub)giant donor
%masses $<2M_\sun$ were dynamically stable

% When the hydrogen shell starts to move into the region with a gradient
% in hydrogen abundace, established during the hydrogen core burning phase,
% and the hydrogen shrinks significantly (Thomas 1967).

\subsection{Limits and influence of $\delta$}
\begin{figure}
\plotone{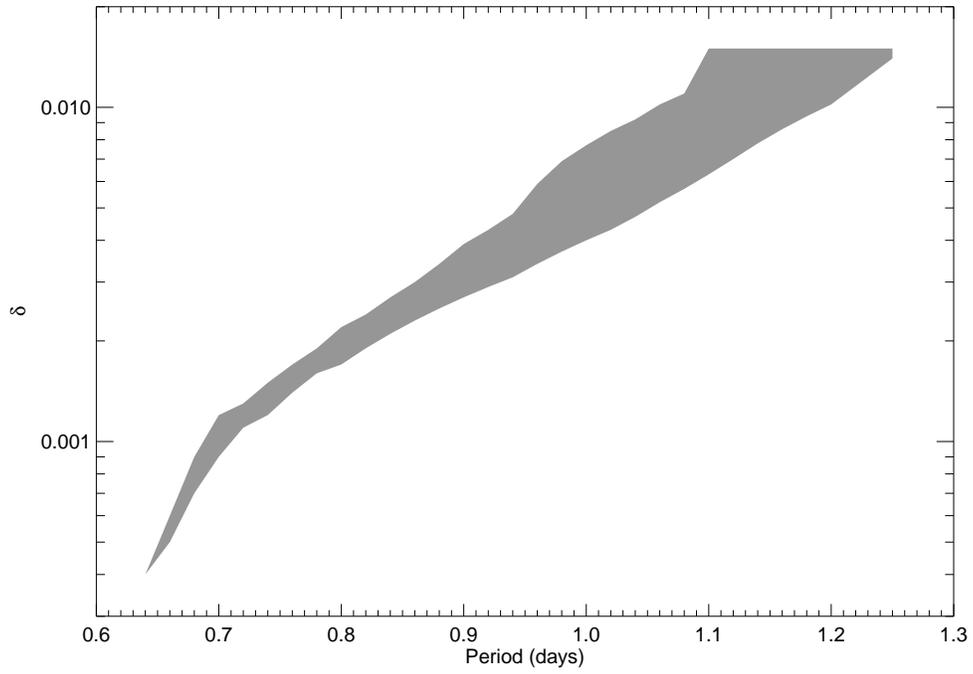} \caption{The distribution of suitable CB disk
parameter $\delta$ and initial orbital period for progenitor binary
systems with $M_{\rm 2,i}=1.1M_\sun$. \label{f2}}
\end{figure}

\begin{figure}
\plotone{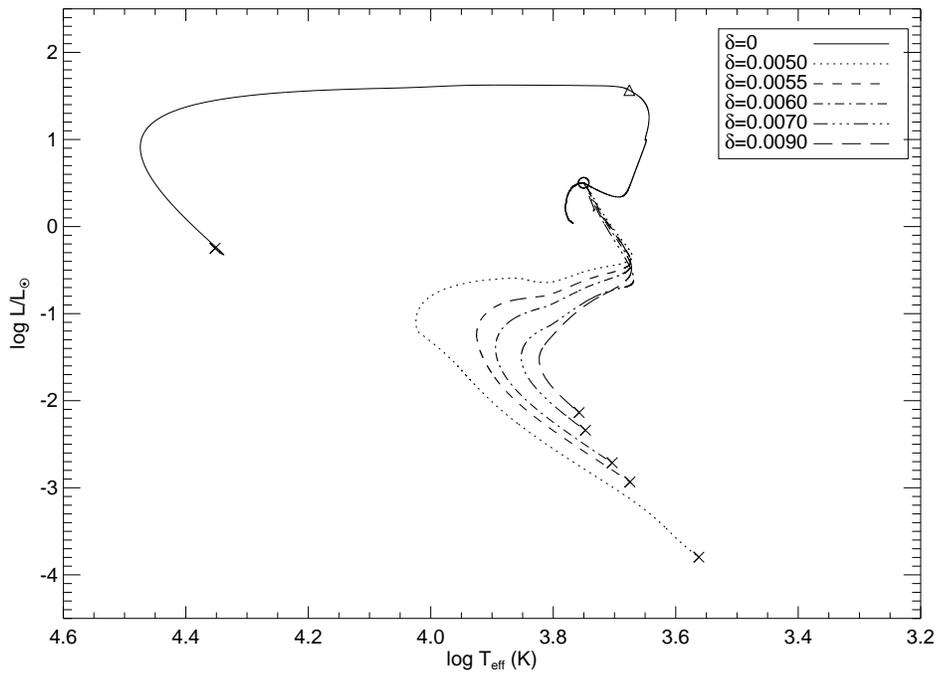} \caption{ Evolutionary tracks of the donor star in
a binary with $M_{\rm 2,i}=1.1M_\sun$, $P_{\rm i}=1.04$ day and
different values of the CB disk parameter $\delta$ in the H-R diagram. The circles
and triangles indicate the beginning and end of mass transfer, respectively. The
crosses correspond to the end of the calculation. \label{HR}}
\end{figure}

\begin{figure}
\plotone{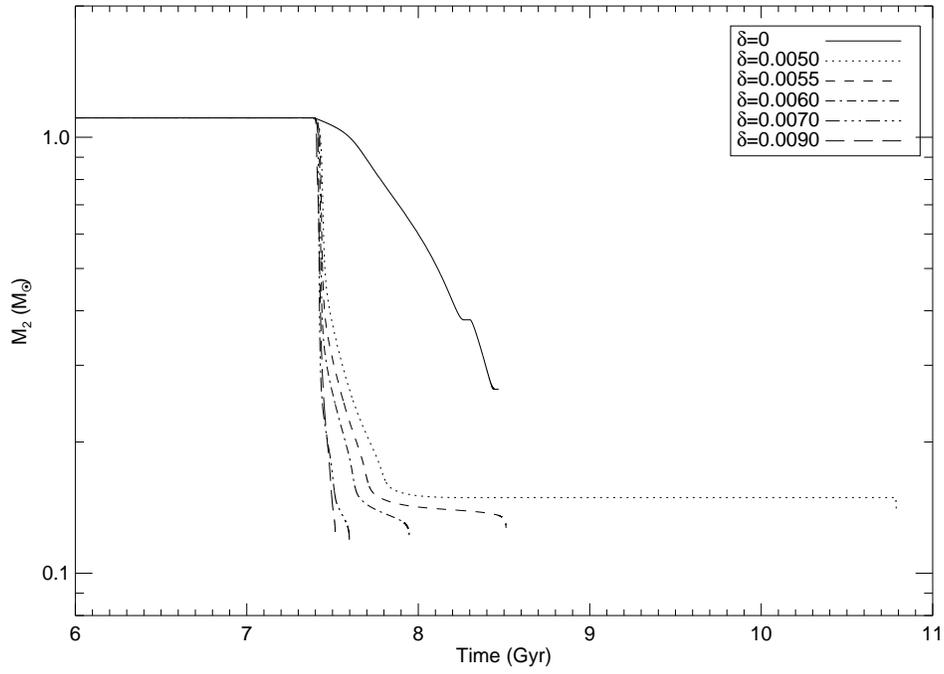} \caption{ Evolution of the donor mass for a binary
with $M_{\rm 2,i}=1.1M_\sun$, $P_{\rm i}=1.04$ day and different values of the CB
disk parameter $\delta$. \label{MP}}
\end{figure}

\begin{figure}
\plotone{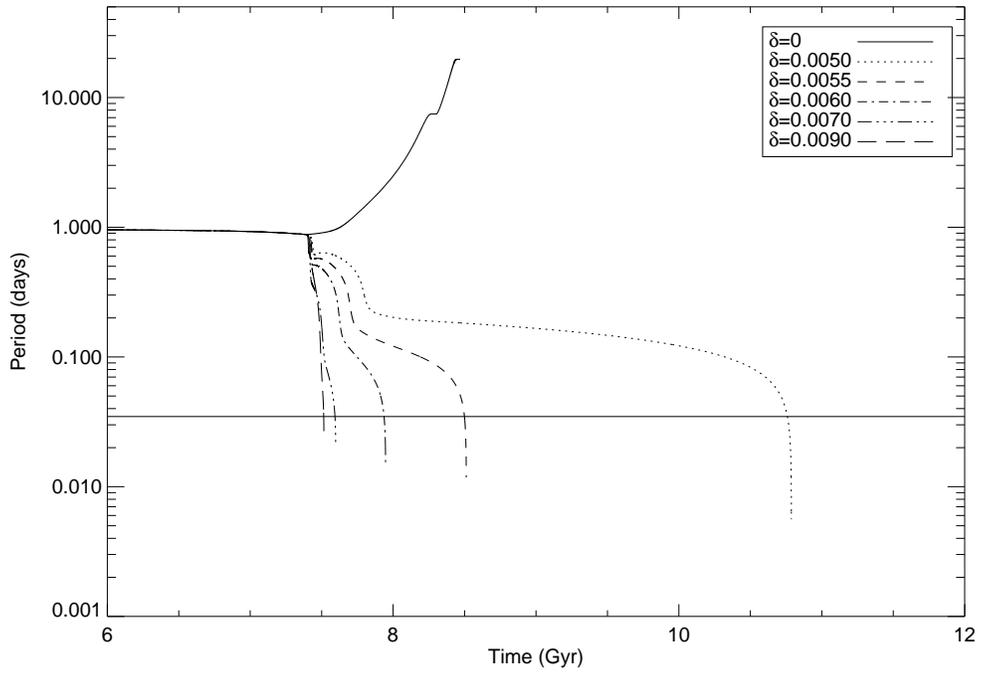} \caption{ Evolution of the orbital period for a
binary with $M_{\rm 2,i}=1.1M_\sun$, $P_{\rm i}=1.04$ day and
different values of the CB disk parameter $\delta$. The horizontal line
corresponds to $P=50$ mins. \label{f5}}
\end{figure}

The possible distribution of $\delta$ and $P_{\rm i}$ with $M_{\rm
2,i}=1.1M_\sun$ is shown in Fig.~\ref{f2}. To illustrate the effects
of $\delta$ on the binary evolution, in Figs.~\ref{HR}$-$\ref{f5},
we plot the exemplarily evolutionary tracks of the secondary in the
H-R diagram, the evolution of  the donor mass and period as a function of
age respectively, for a binary system with
$M_{2,\rm i}=1.1M_\sun$, $P_{\rm i}=1.04$ d and different values of
$\delta$. The lower limit of $\delta$ is determined by the
constraint that age of the binary should be less than $13.7$ Gyr:
a larger value of $\delta$ leads to shorter formation time, as seen from
Fig.~\ref{f5}; if $\delta$ is too small, the binary will not be able
to reach the $50$ min period within $13.7$ Gyr due to inefficient
angular momentum loss. The upper limit of $\delta$ is determined by
two conditions.
%% First condition is the mass transfer has to
%%be stable. For very large $\delta$, the stable mass transfer condition $\xi_{\rm RL} < \xi_{\ ad}$
%%could not be satisfied, thus become dynamical unstable \citep{webbink85}.
The first is that the minimum orbital period $P_{\rm min}$ should be
less than $50$ min. With larger values of $\delta$, the donor will spend
relatively less time on the MS, leaving a smaller degenerate helium
core. According to the mass-radius relation of degenerate stars, the
smaller mass, the larger radius, which corresponds to a larger
$P_{\rm min}$.
The second is $\delta < 0.015$, because we find that mass
transfer becomes unstable in most cases
if $\delta > 0.015$.

The formation and evolutionary paths of UCXBs depend on the adopted
values of $\delta$. In the case of $M_{\rm 2,i}=1.1M_\sun$ and
$P_{\rm i}=1.04$ d, for example, when $\delta<0.0055$, the orbital
period first decreases with mass transfer until the donor star loses
its outer envelope and shrinks rapidly at $P\sim 0.1-0.2$ d. This
causes a cessation of mass transfer. In the subsequent evolution the
orbital period may decrease down to the ultra-short regime under the
effect of GR, until the secondary star fills its RL again, and the
binary appears as a UCXB. When $\delta \ge 0.0055$ the binary
evolves directly into the ultra-short regime with decreasing orbital
period.
%After
%the orbital period reaches its minimum value, it will also bounce
%back because of the degenerate core of the donor star. Thus we could
%possibly observe the period derivative of this binary as positive or
%negative.
%Both of the two channels could explain the negative derivative of the
%$11$ min binary in NGC 6624 \citep{klis93,chou01}.
%( see 9100,110,29, 11100,110,66)
%The mass transfer briefly detaches when shell-burning terminates, but mass transfer
%resumes when shell-burning is well established. (is this right?)

We need to mention that the distribution of $\delta$ depends on the
value of the viscous parameter $\alpha$, which we chose to be
$0.01$ in our calculations as adopted in \citet{spruit01} and
\citet{taam01}.  This is about an order of magnitude lower than the value
($\sim 0.1-0.4$) inferred by \citet{king07}  for fully ionized, thin accretion
disks from observations of dwarf nova outbursts and outbursts
of X-ray transients. However, the $\alpha$ value estimated by \citet{king07} 
is an average one over the entire disk, while here it is at the
inner edge of the disk \citep{spruit01}, which may be smaller
due to the boundary condition \citep[e.g.][]{pap03,winters03,fromang06}. 
Additionally, since the CB disk is located
outside the binary, and hence shielded from X-ray irradiation from the NS 
by the accretion disk around the NS and the secondary, 
the $\alpha$ value in a CB disk might also be smaller than in accretion disks.
Taking account of the above facts, we suggest to regard Fig.~\ref{f2}  as 
an optimistic case for the distribution of $\delta$.
Nevertheless, from Eqs.~(3) one can see that the CB disk-induced angular
momentum loss rate is proportional to
$\alpha^{1/3}\delta$. So if keeping $\alpha^{1/3}\delta$ 
constant, the binary evolution will be exactly the same (this has been
verified by our test calculations), implying a predictable 
$\delta$ distribution for a given value of $\alpha$.

\subsection{Comparison with observations}

\begin{figure}
\plotone{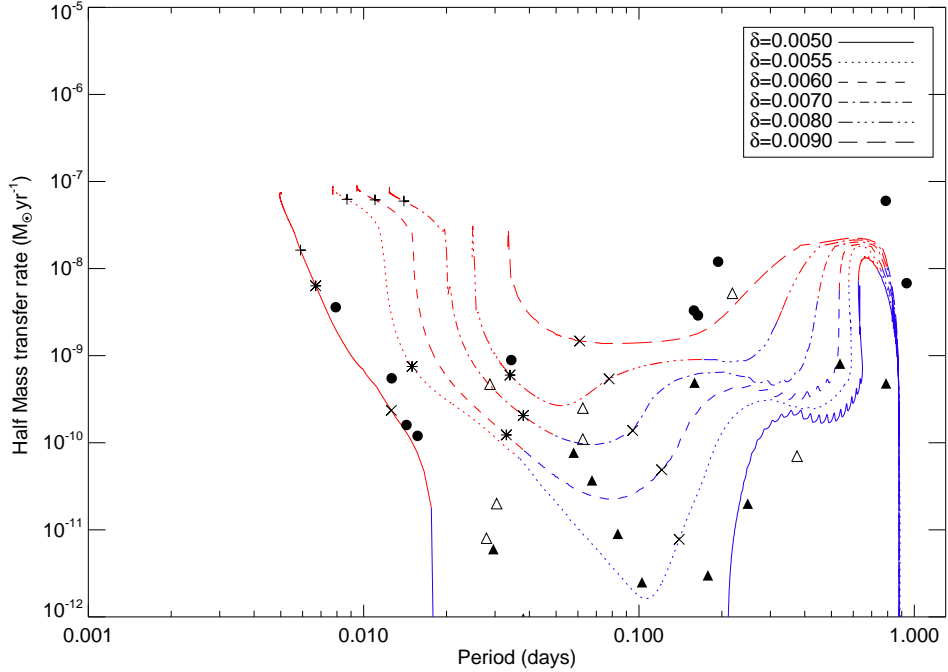} \caption{Evolution of mass accretion rate (or half of the mass
transfer
rate) vs. orbital period for binary systems with $M_2=1.1M_\sun$,
$P_{\rm i}=1.04$ day and different values of the CB disk parameter $\delta$.  The
red and blue lines indicate persistent and transient accretion
according to the criteria in \citet{lasota08} for a
mixed-composition ($X=0.1$, $Y=0.9$) disk. Persistent (filled
circles) and transient (triangles) LMXBs (including UCXBs) are also
plotted for comparison. Here the open triangles mean that the derived mass
accretion rates from observations are the upper limits. The symbols $\times, \ast, +$
on the evolutionary sequences denote where the composition of the donor
in the binary is $X=0.3$, 0.2, and 0.1 with $Y=0.98-X$ and $Z=0.02$, respectively.
%( $+$: 4U 1820$-$30, $\ast$: 4U 1543$-$624, $\times$: 4U 1850$-$087, $\square$: M15 X-2,
%$\vartriangle$: 4U 1916$-$05).
\label{f6}}
\end{figure}

\begin{figure}
\plotone{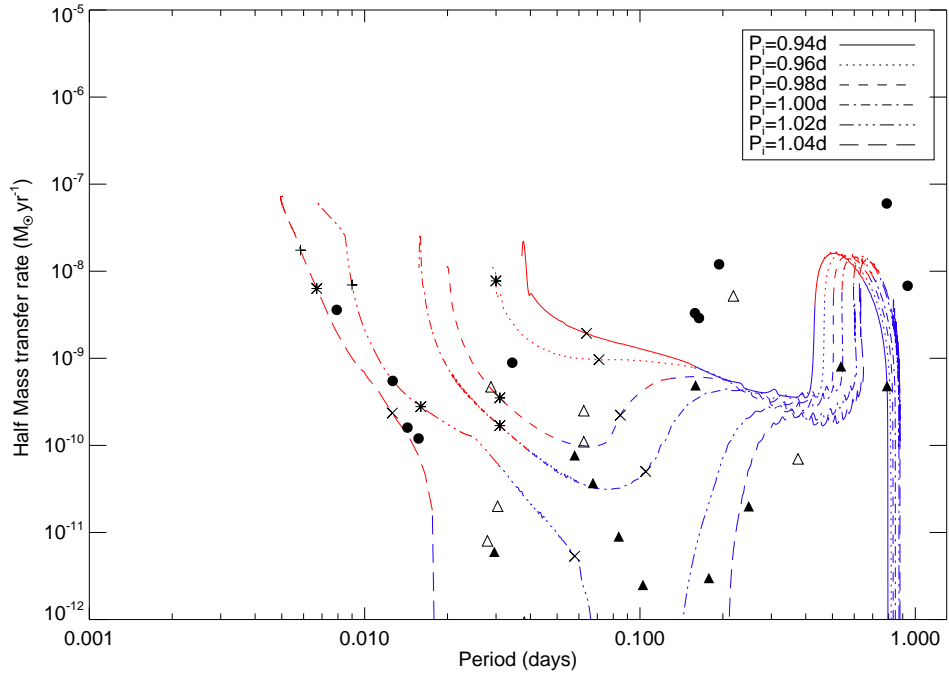} \caption{Same as in Fig.~6 but for
binary systems with $M_{2, {\rm i}}=1.1M_\sun$,
$\delta=5\times10^{-3}$, and $P_{\rm
i}=0.94-1.04$ day.
%Periods and mass transfer rates of 5 persistent UCXBs are also marked in this figure
%( $+$: 4U 1820$-$30, $\ast$: 4U 1543$-$624, $\times$: 4U 1850$-$087, $\square$: M15 X-2,
%$\vartriangle$: 4U 1916$-$05).
\label{f7}}
\end{figure}

There are currently 10 UCXBs with known periods, 5 of which are
persistent sources and 5 are transients. We list the orbital periods
$P_{\rm orb}$ and mean mass accreting rates $\dot{M}_1$ (or the
upper limit of $\dot{M}_1$) of these UCXBs in Table~1. The
$\dot{M}_1$s listed in this table for persistent UCXBs are
calculated by using the luminosities mentioned in the references and
assuming accretion onto a $1.4M_\sun$ NS with a $10$ km radius,
while those for transient sources are from the estimates of
\citet{krauss07}, \citet{watts08} and \citet{lasota08}. To compare
observations with our CB disk-assisted binary model, we plot the
$\dot{M}_1(=-\dot{M}_2/2)$ vs. $P_{\rm orb}$ relations in Fig.~\ref{f6} for
binary systems with $M_2=1.1M_\sun$, $P_{\rm i}=1.04$ d and
$\delta=0.005-0.009$, and in Fig.~\ref{f7} for binary systems with
$M_2=1.1M_\sun$, $\delta=0.005$ and $P_{\rm i}=0.94 -1.04$ d.
Note that in these two figures the values of $\dot{M}_1$ are not assumed
to be limited to $\dot{M}_{\rm Edd}$,
so that possible super-Eddington accretion can be allowed. In this
way we can compare our results with observations directly. We also
indicate in Figs.~\ref{f6} and \ref{f7} whether the accretion disks in
the LMXBs are thermally and viscously stable, according to the
stability criterion for a mixed-composition ($X=0.1$, $Y=0.9$) disk
from \citet{lasota08}
\footnote{We notice two points
about the persistent/transient criterion in UCXBs. Firstly, in
\citet{lasota08} it is found that three of the five persistent UCXBs
should be transient if their accretion disks are composed of pure
helium or elements heavier than helium (C/O). However, in our CB
disk model, the disks are not composed of pure helium but of
mixed-compositions, so the accretion disks in these UCXBs are
thermally stable, consistent with observations. Secondly, the
transient source 4U1626$-$67 should be persistent according to the
criterion, but it is transient in a different way: its outbursts do
not last tens of days but tens of years \citep{krauss07}. The value
of $\dot{M}_1$ in Table 1 is calculated from Eqs.~(4) in
\citet{krauss07} assuming a distance of 3 kpc \citep{chakrabarty98}
and the time between outbursts to be 30 years. If the recurrence
time is as long as 1000 yrs, it will yield a much lower $\dot{M}_1$
\citep{lasota08}, and may help to resolve this transient/persistent
problem. }.
We use the symbols $\times$, $\ast$, and $+$ on the
evolutionary tracks to denote where the hydrogen composition $X$
of the donor becomes 0.3, 0.2, and 0.1, respectively,
to show that the criterion of \citet{lasota08} is applicable here.
%As we have
%mentioned in \S 3.3, there are two formation channels in this model.
%The binary models with $\delta=0.005$ and $P_{\rm i}=1.02, 1.04$ day
%in these two figures are in channel 1, while the others are in
%channel 2.
The positions of UCXBs are marked in these two
figures with circles and triangles for persistent and transient sources,
respectively. Besides them, we also include 18 NS LMXBs
with known $P$ and $\dot{M}_1$
\citep[data are taken from ][]{liu07,watts08,heinke09}.

A comparison between our CB disk-assisted binary models and the
observations of (compact) NS LMXBs suggesting
that it is possible to form UCXBs from normal LMXBs. We note that
three of the UCXBs are in globular clusters, indicating low
metallicities in these systems. However, from our calculations we
find that change of metallicities does not significantly affect the
binary evolution when the CB disk is involved.
The 11-min UCXB 4U 1820$-$30 is particularly interesting because
of its negative period derivative $\dot{P}/P=-3.5\pm1.5\times
10^{-8}$ yr$^{-1}$ \citep{klis93a,chou01}, which is inconsistent
with the lower limit ($\dot{P}/P>8.8\times 10^{-8}$ yr$^{-1}$) of
the standard evolutionary scenario \citep{rapp87}. While previous
explanations for this negative period derivative invoke acceleration
of the binary by a distant third companion in a hierarchical triple
system, or by the cluster potential
\citep{tan91,klis93b,king93,chou01}, our CB disk scenario may  present
an alternative interpretation
of this negative period derivative (see Fig.~5).

\section{Discussion and Conclusion}
During RLOF mass transfer, a CB disk may be formed as a result of mass outflow from the
accretion disk, and has been invoked as an efficient process for
the removal of orbital angular momentum
\citep{taam01}. We propose a scenario for the formation of UCXBs
from L/IMXBs with the aid of a CB disk in this work. The suitable
binary parameter space ($M_{\rm 2,i}$ and $P_{\rm i}$) with
reasonable choice of the CB disk parameter $\delta$ for the
formation of UCXBs within $13.7$ Gyr is found to be significantly
larger than in previous ``magnetic capture" model
\citep{sluys05a,sluys05b}. This difference is caused by the fact
that the bifurcation period is considerably increased if the CB disk
is included. In L/IMXB evolution the bifurcation period $P_{\rm
bif}$ is defined as the initial orbital period when the donor star
is on ZAMS, which separates the formation of converging systems from
diverging systems.  Because the value of $P_{\rm bif}$ depends
strongly on the angular momentum loss mechanisms
\citep{sluys05a,ma08}, we would expect $P_{\rm bif}$ to be
significantly changed when the CB disk is included in the binary model.
In Table~2 we present the calculated values of $P_{\rm bif}$ for
binary systems consisting of a $1.4\,M_\sun$ NS and a $1.1M_\sun$
secondary, with ($\delta=0.002-0.01$) and without a CB disk. We also
list the corresponding values of the period ($P_{\rm rlof}$) at which RLOF begins
\citep{pod02,ma08}. According to the investigation of \citet{pod02}
and \citet{sluys05a}, binary systems with initial
orbital period slightly below the bifurcation period can achieve
the shortest possible orbital period. We list the shortest periods
for binary systems with certain value of $\delta$ within $13.7$ Gyr
in Table 2, which clearly indicate that, (1) the larger $\delta$,
the larger $P_{\rm bif}$ and $P_{\rm rlof}$, and (2) UCXBs ($P<50$
min) are not likely to form in such a scenario without the aid of CB disks
($\delta=0$).

We note here that when the binaries reach their minimum periods
where the donors become degenerate, their orbital periods will
bounce back into the period-increasing phase. Our code cannot follow
the binary evolution with a degenerate donor star. It is likely that
the observed UCXBs may be explained as binaries both evolving to
shorter orbital periods with a hydrogen-deficient,  non-degenerate
donor star (as presented in Figs.~6 and 7), and with a degenerate WD
donor during a period-increasing phase
\citep{yungelson02,nelson03,deloye03,deloye07}. In addition, the
helium-donor channel (or semi-degenerate channel) may also
contribute to the formation of UCXBs \citep{savonije86}. Currently
it is difficult to compare the (spectral) theoretical models for
hydrogen-deficient accretion disks with observations, or directly
measure the orbital period derivative, one possible criterion to
discriminate  UCXBs  in the period-decreasing/increasing phases is
related to the donor masses, which, together with the orbital
periods,  allow to determine the mass-radius relation of the donor.
Our calculations show that the UCXB's donor mass is around $\sim 0.1
M_\sun$ in the period-decreasing phase at $P\sim 40$ mins, while in
the period increasing phase, the donor mass should be around $0.01
M_\sun$ \citep[e.g.][]{yungelson02,nelson03}. Previously
investigations on the UCXBs XTE J1807$-$294 \citep{falanga05}, SWIFT
J1756.9$-$2508 \citep{krimm07}, and XTE J0929$-$314
\citep{galloway02} have shown that the donors should be WDs unless
the binary orbital inclination is very small ($<10\degr$), because
of their small mass functions, while XTE 1751$-$305 is more likely
to be in the period-decreasing phase due to its relatively larger
mass function \citep{markwardt02}. Observationally there seem to be
more systems in the period-increasing phase than in the
period-decreasing phase. This may be addressed by the fact that
UCXBs spend longer mass-transfer time in the former \citep[$>10^8$
yr, see][]{rasio00,deloye03} than in the latter phase \citep[$\sim
10^7$ yr, see][and this wrok]{nelson03}. From Table~1 there appears
to be an apparent accumulation of systems with $P\sim 40-50$ mins.
The reason may lie in that (1) during the period-increasing phase,
UCXBs spend more time at larger orbital periods
\citep[see][Fig.~9]{deloye03}, and (2) when $P> 50 $ mins, most of
these systems become transient sources with very weak accretion
($<10^{-12}M_\sun$yr$^{-1}$). From the calculation of
\citet{deloye05}, these semi-degenerate systems should mainly
distribute at $40$ mins $ <P<90$ mins \citep[see][Fig.~5]{deloye05}.
Most of them should be transients \citep[see][Fig.~2]{deloye05},
among which the longer the orbital period, the weaker the accretion
will be, thus the more difficult to be detected. These together
account for the accumulation of UCXBs with $P\sim 40-50$ mins.
Obviously a thorough population synthesis is needed to address the
contribution to UCXBs from systems with non-degenerate,
semi-degenerate and degenerate donors.

However, there exist some issues in the CB disk scenario, especially
the existence of CB disk in LMXBs. \citet{dubus02} suggest that CB
disks are prospective to be observed in infrared and sub-millimeter
band. Although observations have shown evidence for the existence of
CB disks in young binary systems \citep{monnier08, ireland08},
magnetic cataclysmic variables
\citep{howell06,brinkworth07,dubus07,hoard07}, and black hole LMXBs
\citep{muno06,gallo07}, more infrared observations are still needed
to confirm or disprove the hypothesis that CB disk may exist in some
LMXBs \citep[e.g.][]{dubus04}. Additionally, the CB disk parameter
$\delta$ is poorly known, and it is possible to change with time or
mass transfer rate. The strong dependence of LMXB evolution  on the
value of $\delta$ prevents accurate estimation of the contribution
of such binaries to UCXBs.  More generally, we do not insist that
there should be a CB disk in L/IMXBs, but argue that a mechanism
that mimics its features may be an important ingredient for
understanding the overall period distribution of UCXBs as well as
cataclysmic variable binaries \citep{willems05}.

\begin{figure}
\plotone{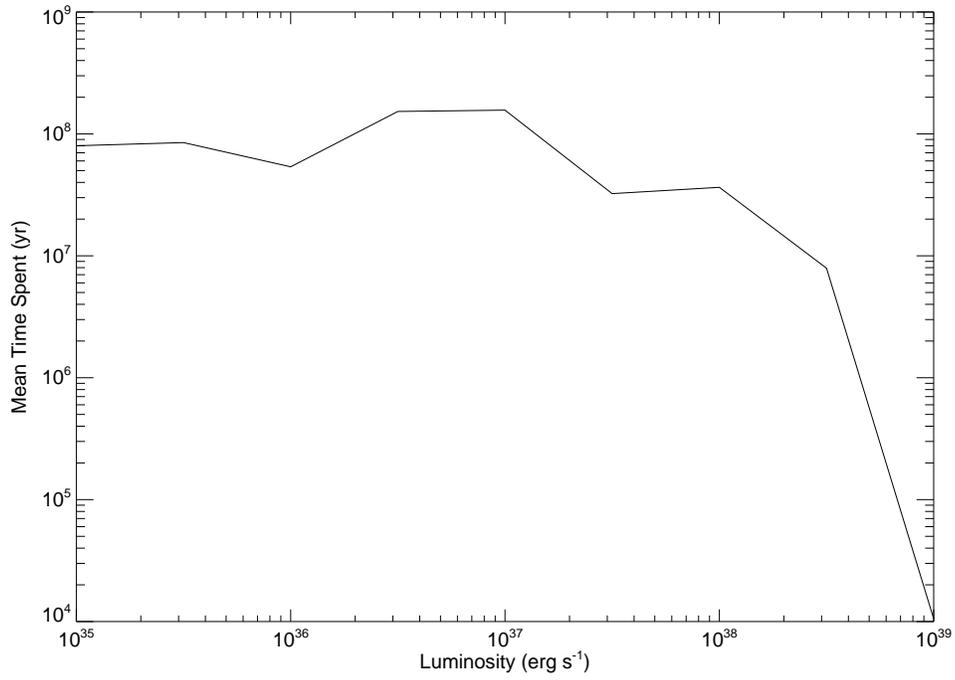} \caption{Mean X-ray lifetime spent by the LMXBs
which could evolve to UCXBs within $13.6$ Gyr at certain
luminosities with $\delta=0.005$. The mean life time decreases
sharply when $L>3\times10^{38}$ ergs$^{-1}$. \label{f8}}
\end{figure}

Recent {\em Chandra} observations of nearby elliptical galaxies have
revealed a population of luminous point X-ray sources, which are
likely to be LMXBs with accretion rates $\dot{M}>10^{-8} M_\sun$
yr$^{-1}$ \citep[e.g.][]{gilfanov04,kim04}. These sources are
explained either as transient LMXBs in which NSs accreting from a
red-giant star in wide orbits ($P>10$ d) \citep{piro02} or
ultra-compact binaries ($P\sim 8-10$ min) with a $0.06-0.08M_\sun$
He or C/O donor \citep{bildsten04}. Our calculations suggest that
normal LMXBs with a CB disk may present a plausible alternative
interpretation for these luminous X-ray sources. We show in
Fig.~\ref{f8} the mean lifetime spent by LMXBs evolved to UCXBs at
certain luminosity with $\delta=0.005$. Here by the luminosity we
mean the ``potential maximum luminosity", where the Eddington limit
is removed and nearly all the mass lost by the donor is assumed to be
accreted by the NS. From this figure we see that the UCXBs can be
luminous ($L>10^{38}$ ergs$^{-1}$) for $\sim 10^7$ yr, and the X-ray
lifetime decreases sharply when
$L>3-5\times10^{38}$ ergs$^{-1}$, which may account for the break in
the luminosity function at $\sim 5\times10^{38}$ ergs$^{-1}$ \citep{kim04}. A
distinct feature of this explanation is that the luminous X-ray
sources are predicted to be short-period, persistent rather transient
sources.

%We give a third explanation here. In our calculation we find that the CB disk could
%significantly increase the loss of angular momentum, thus could be able
%to raise the mass transfer rate to the Eddington limit at $P_{\rm orb}=5-80$
%min (see also Fig.~4 and Fig.~5). If we could truly detect the orbital
%periods of these point sources in the range of $10-80$ min, they could
%not be explained by the second explanation above, but could be accounted
%as an accreting NS driven by a CB disk. So the best confirmation of these
%conjectures is to detect the periodicity of these sources.

\begin{acknowledgements}
We thank an anonymous referee for his/her valuable comments that helped
improve the original manuscript. BM thanks W.-C. Chen and P.  P.
Eggleton for helpful discussions and suggestions. This work was
supported by Natural Science Foundation of China under grant
10873008 and National Basic Research Program of China (973 Program
2009CB824800).

\end{acknowledgements}

\clearpage

\clearpage

%\begin{figure}
%\plotone{f5.eps}
%\caption{ Evolution of the orbital period for binary systems with $M_2=1.1M_\sun$,
%$\delta=5\times10^{-3}$ with initial orbital period $P_{\rm i}= 0.94-1.04$ day.
%\label{f5}}
%\end{figure}

\clearpage

\begin{table}
\begin{center}
\caption{Parameters of known UCXBs}
\begin{tabular}{c  c  c c c}
\tableline \tableline
Systems & Type & $P_{\rm orb}$ (min) & $<\dot{M}>(M_\sun$y$^{-1})$ & Refs \\
\tableline
4U 1820-30 & P & 11.4 & $3.6^{+1.5}_{-0.6}\times 10^{-9}$ & 1\\
4U 1543-624 & P & 18.2 & $5.5^{+40}_{-4}\times 10^{-10}$ & 2\\
4U 1850-087 & P &20.6 & $1.6\pm0.3\times 10^{-10}$ & 3\\
M15 X-2  & P &  22.6 & $1.2\pm0.2\times10^{-10}$ & 4\\
4U 1916-05 & P & 49.5 & $8.9\pm1.3\times10^{-10}$ & 5 \\
XTE J1807-294 & T & 40.07 & $<8\pm7\times10^{-12}$ & 6,10\\
XTE J1751-305 & T & 42.42 & $6\pm5\times10^{-12}$ & 6\\
XTE J0929-314 & T & 43.6 & $<2\pm1.5\times10^{-11}$ & 6,11\\
4U 1626-67  & T & 41.4 & $4.7^{+5.1}_{-3.2}\times10^{-10}$ & 7,8\\
SWIFT J1756.9-2508 & T & 54.7 &$9.3\pm7\times10^{-13}$ &9\\
\tableline
\end{tabular}
\tablecomments{Here the type means persistent (P) or transient (T)
UCXBs. The mean mass transfer rates $<\dot{M}>$ listed here for
persistent UCXBs are calculated by using the luminosities mentioned
in the references and assuming accretion onto a $1.4M_\sun$ NS with
$10$ km radius. The uncertainties in the calculated $<\dot{M}>$
mainly come from the uncertainties of source distance, neutron star
masses and radii, spectral fit goodness, and
recurrence times of transient sources (see \citet{heinke07, heinke09}
for a thorough discussion for these factors). }
\tablerefs{ (1) \citet{zd07}, \citet{kuulkers03},
\citet{shaposhnikov04}; (2) \citet{wang04}, \citet{schultz03}; (3)
\citet{sidoli06}, \citet{harris96}, \citet{pal01}; (4)
\citet{dieball05}, \citet{hannikainen05}; (5) \citet{juett06}; (6)
\citet{heinke09}; (7) \citet{chakrabarty98}; (8) \citet{krauss07};
(9) \citet{lasota08}, \citet{krimm07}; (10) \citet{falanga05}; (11)
\citet{wijnands05}. }
\end{center}
\end{table}

\clearpage

\begin{table}
\begin{center}
\caption{The bifurcation periods $P_{\rm bif}$ and $P_{\rm rlof}$, and the
shortest periods $P_{\rm min}$ achieved for a binary with $M_{\rm
2,i}=1.1M_\sun$ and different values of $\delta$.}
\begin{tabular}{c  c  c c}
\tableline \tableline
$\delta$ & $P_{\rm bif}$ (day) & $P_{\rm rlof}$ (day) & $P_{\rm min}$ (min)\\
\tableline
0     & 0.63 & 0.5  & 71\\
0.002 & 0.87 & 0.75 &  9\\
0.004 & 1.08 & 0.92 &  7\\
0.006 & 1.15 & 0.96 &  6\\
0.008 & 1.21 & 1.04 &  6\\
0.010 & 1.30 & 1.13 &  6\\
\tableline
\end{tabular}
%\tablecomments{Please see the text for the meaning of $P_{\rm bif}$,
%$P_{\rm rlof}$, $P_{\rm min}$}
%\tablerefs{}
\end{center}
\end{table}

\end{document}